\documentclass[a4paper,11pt]{article}
\usepackage{pos}
\usepackage{caption}
\usepackage{subcaption}

\usepackage{amsmath}
\usepackage{dsfont}

\usepackage{graphicx}
\title{The deconfinement phase transition in $\SpN$ gauge theories and the density of states method}
\ShortTitle{Density of states in $\SpN$ gauge theories}

\author*[a]{David Mason}
\author[b,c]{Biagio Lucini}
\author[a]{Maurizio Piai}
\author[d]{Enrico Rinaldi}
\author[e]{Davide Vadacchino}

\affiliation[a]{Department of Physics, Faculty of Science and Engineering, Swansea University (Park Campus),
Singleton Park, SA2 8PP Swansea, Wales, United Kingdom}
\affiliation[b]{Department of Mathematics, Faculty of Science and Engineering, Swansea University (Bay Campus),
Fabian Way, SA1 8EN Swansea, Wales, United Kingdom}
\affiliation[c]{Swansea Academy of Advanced Computing, Swansea University (Bay Campus), Fabian Way, SA1 8EN Swansea, Wales, United Kingdom}
\affiliation[d]{Interdisciplinary Theoretical \& Mathematical Science Program, RIKEN (iTHEMS), 2-1 Hirosawa, Wako, Saitama, 351-0198, Japan}
\affiliation[e]{Centre for Mathematical Science, University of Plymouth, Plymouth, PL4 8AA, United Kingdom}

\emailAdd{2036508@Swansea.ac.uk}
\emailAdd{b.lucini@swansea.ac.uk}
\emailAdd{m.piai@swansea.ac.uk}
\emailAdd{erinaldi.work@gmail.com}
\emailAdd{davide.vadacchino@plymouth.ac.uk}

\abstract{First-order phase transitions in the early universe might produce a detectable background of gravitational waves. As these phase transitions can be generated by new physics, it is important to quantify these effects. Many pure Yang-Mills gauge theories are known to undergo first-order deconfinement phase transitions, with properties that can be studied with lattice simulations. Despite the recent surge of interest in $\SpN$ gauge theories as a candidate for models of physics beyond the standard model, studies of these theories at finite temperature are still very limited. 
In this contribution we will present preliminary results of an ongoing numerical investigation of the thermodynamic properties of the deconfinement phase transition in $\SpF$ Yang-Mills theory, using the linear logarithmic relaxation algorithm. This method enables us to obtain a highly accurate determination of the density of states, allowing for a precise reconstruction of thermodynamic observables. In particular, it gives access to otherwise difficult to determine quantities such as the free energy of the system, even along metastable and unstable branches, hence providing an additional direct observable to study the dynamics of the phase transition.}

\FullConference{The 40th International Symposium on Lattice Field Theory (Lattice 2023)\\
July 31st - August 4th, 2023\\
Fermi National Accelerator Laboratory\\}
\newcommand{\beq}{\begin{equation}}
\newcommand{\eeq}{\end{equation}}

\begin{document}

\newcommand{\sun}{$SU(N_c)$~}
\newcommand{\suTN}{$SU(2N)$~}
\newcommand{\suthree}{$SU(3)$~}

\makeatletter
\newsavebox{\@brx}
\newcommand{\llangle}[1][]{\savebox{\@brx}{\(\m@th{#1\langle}\)}%
  \mathopen{\copy\@brx\mkern2mu\kern-0.9\wd\@brx\usebox{\@brx}}}
\newcommand{\rrangle}[1][]{\savebox{\@brx}{\(\m@th{#1\rangle}\)}%
  \mathclose{\copy\@brx\mkern2mu\kern-0.9\wd\@brx\usebox{\@brx}}}
\newcommand{\SpN}{S\mspace{-2mu}p(2N)}%
\newcommand{\SpF}{S\mspace{-2mu}p(4)}%
\makeatother

\maketitle
\section{Introduction}
\label{intro}
New strongly interacting sectors based on symplectic gauge theories have properties which are phenomenologically interesting to physics beyond the standard model. 
For a recent review on this see Ref.~\cite{Bennett:2023wjw} and references therein. 
Due to the non-perturbative nature of these theories, the lattice is the natural first-principles method to study their properties.

If, as is the case for deconfinement in $\SpN$ $(N>1)$, these theories undergo first-order phase transitions in the early universe, they can leave imprint through the generation of long wavelength gravitational waves, opening gravitational wave astronomy as a promising route to constrain new physics, see for instance discussion from Ref.~\cite{Pasechnik:2023hwv} and references therein. 
The recent announced detection of a gravitational wave background from NANOGrav would be consistent with this type of signal~\cite{NANOGrav:2023hvm}. 
However, to constrain new physics, accurate estimations of the expected power spectrum from theory is required. 
The lattice literature on $\SpN$ gauge theories at finite temperature is quite limited~\cite{Holland:2003kg, BrunoInPreparation}. 
We present the initial results of a project which aims to accurately calculate the thermodynamic properties of the deconfinement phase transition in $\SpN$ gauge theories through the use of a novel lattice method, the logarithmic linear relaxation (LLR) method~\cite{Lucini:2023irm, Mason:2022trc,Mason:2022aka, Langfeld:2012ah, Langfeld:2013xbf, Langfeld:2015fua, Cossu:2021bgn, Springer:2021liy, Springer:2022qos, Springer:2023wok, Lucini:2016fid}.

Two important quantities of interest are latent heat and surface tension. 
In the critical region the system will exhibit co-existing phases separated by a potential barrier.
The probability of tunnelling through the barrier is related to the surface tension and the latent heat is equal to the energy difference between the phases.
Through continuum and infinite volume extrapolations of observables of the lattice system, these quantities can be computed in the continuum theory. 

Sec.~\ref{sec:Setup} presents the lattice setup and details of the LLR algorithm. In Sec.~\ref{sec:results}, we report the thermodynamic observables for a single lattice size, $N_t \times N_s^3=4\times20^3$, for $\SpF$ pure gauge theory, and we discuss and compare our results to those of $\suthree$ pure gauge theory~\cite{Lucini:2023irm} --- see also the comprehensive discussion in Ref.~\cite{Borsanyi:2022xml}.  
\section{Lattice setup and the LLR method}
\label{sec:Setup}
We use an $N_t\times N_s^3$ hypercubic lattice in Euclidean spacetime, with periodic boundary conditions, spacing $a$ and volume $\tilde{V}=a^4N_t N_s^3$. When $N_t<N_s$, the temperature of this system is given by the inverse of the temporal extent $T=1/aN_t$. We use the standard Wilson action, defined by
\beq
\label{eqn:Action}
S[U]\equiv\frac{6\tilde{V}}{a^4}(1-u_p[U]),
\eeq
where $U$ refers to the lattice configuration and $u_p$ is the average plaquette. The partition function of this system at coupling $\beta$ is given by
\beq
\label{eqn:PathIntegral}
Z_\beta \equiv  \int [DU_\mu]e^{-\beta S[U]}.
\eeq
To study the thermodynamic properties across the phase transition, the temperature is varied by keeping the number of sites fixed and altering the lattice spacing through the coupling, $\beta(a)$. 

The lattice configurations, $U$, consist of link variables $U = \{U_\mu(n_t, \vec{n_s})\}$, where $U_\mu \in \SpN \subset \suTN$, obey the condition $U_\mu \Omega (U_\mu)^T = \Omega$.
The symplectic matrix, $\Omega$, is defined as 
\beq
\Omega = \begin{pmatrix}
0 & \mathds{1}_{N\times N}\\
-\mathds{1}_{N\times N} & 0
\end{pmatrix}.
\eeq

The centre of the group $\SpN$ is $\mathbb{Z}_2$, for all finite $N$. On the lattice, the deconfinement phase transition is associated with the spontaneous breaking of the centre symmetry. The corresponding order parameter is the average Polyakov loop, defined as
\beq
\label{eqn:PolyakovLoop}
\left\langle  l_p  \right\rangle _\beta \equiv \left\langle \frac{1}{2 N N_s^3} \sum_{\vec{n _s}} \textrm{Tr}\left(\prod_{n_t = 0}^{N_t - 1} U_0(n_t, \vec{n _s})\right)  \right\rangle_\beta  \ \  \begin{cases} = 0 \text{ confined phase} \\
\neq 0 \text{ deconfined phase}
\end{cases} \ .
\eeq

For $N>1$, $\SpN$ gauge theories have first-order deconfinement phase transitions. 
The co-existence of phases in the proximity of first-order transitions leads to metastable dynamics which is problematic when using Monte Carlo importance sampling methods. 
To accurately sample the phase space, the system must tunnel between the vacua many times. 
For small lattices, in general this is not a problem as the potential barrier is small. 
However, as the lattice volume increases, the potential barrier grows and more configurations are required to obtain accurate results. 

Following the results presented in Ref.~\cite{Lucini:2023irm}, we employ the logarithmic linear relaxation method to overcome the metastability problems around the deconfinement phase transition. 
The goal of the LLR method is to accurately estimate the density of states,
\beq
\rho(E) \equiv \int [DU] \delta (S[U] - E),
\eeq
as a piecewise log linear function, with the total energy range relevant to the processes of interest broken down into $2N_I-1$ intervals,  
\beq
\label{eqn:LLRRho}
\ln \rho (E) \approx a_n (E-E_n) + c_n. 
\eeq 
where $n=1,...,2N_I - 1$, denotes the energy interval that the expression is valid for, $E_n - \Delta_E/4 \leq E \leq E_n + \Delta_E/4$. The $c_n$ term is set by the continuity of $\rho$ at the boundary of the intervals, $c_n = a_1\Delta_E/4 + (\Delta_E/2)\sum_{k=2}^{n-1}a_k + \Delta_E a_n /4$. We take $E_1 \neq 0$, therefore $c_1$ is arbitrary and we set it to 0. 

In each interval we calculate the coefficient, $a_n$, by solving the equation $\langle \langle E-E_n \rangle \rangle_n(a_n) = \langle \langle u_p-(u_p)_n \rangle \rangle_n(a_n) = 0$, where $(u_p)_n = 1 - E_n a^4 /6 \tilde V$. The double angle bracket  $\langle \langle ...\rangle \rangle_n(a_n)$ denotes the expectation value for configurations restricted to the $n^{th}$ energy interval, for a coupling $a_n$. In this work, these expectation values are computed through a modified heat bath algorithm, as discussed in Ref.~\cite{Lucini:2023irm}.

We solve these equations, to find the set of $a_n$ values, through a combination of Newton-Raphson and Robbins-Monro iterations, Eq.~(12) and (14)  of Ref.~\cite{Lucini:2023irm} respectively. 
In the limit of infinite Robbins-Monro iterations, the exact solution would be found, however only a finite number of iterations are possible, introducing a truncation error.
The truncation error is estimated by repeating the determination of the $a_n$ values and bootstrapping the results. 

Once $\{a_n\}_{n=1}^{2N_I-1}$ has been found, the density of states can be reconstructed. This is used to compute the plaquette, or equivalently energy, distribution of the system at coupling $\beta$, through the relation
\beq
\label{eqn:PlaqDistribution}
P_{\beta}(u_p)  = \frac{1}{Z_\beta}\rho(E)e^{-\beta E}|_{E=6\tilde V(1-u_p)/a^4}, \quad Z_\beta= \int dE \rho(E) e^{-\beta E}.
\eeq
Expectation values can then be reconstructed avoiding the metastability problem.

\begin{figure}
\centering
\includegraphics[width=0.6\textwidth]{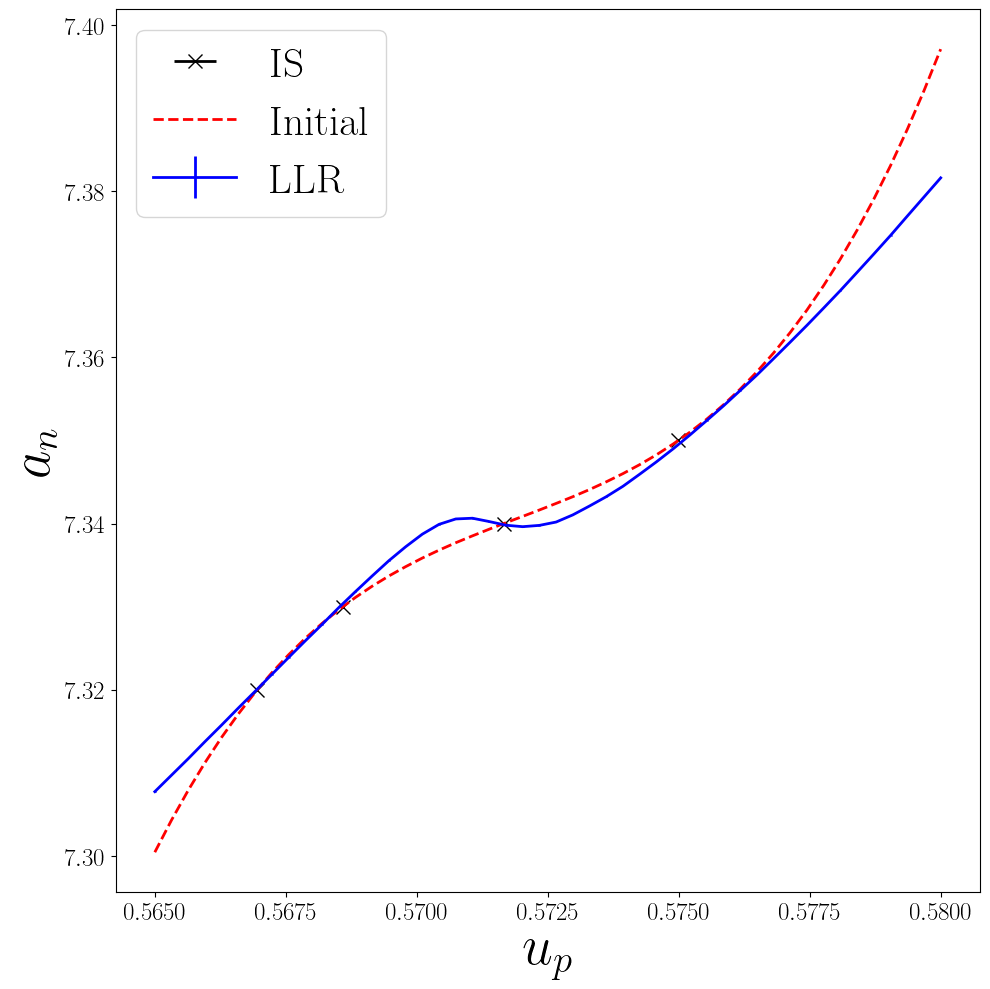}
\caption{The values of $a_n$ against $(u_p)_n=1-E_n a^4 / 6\tilde{V}$, the plaquette values corresponding to the centre of each energy interval, for $\SpF$ pure gauge theory on a lattice of size $4 \times 20^3$. We use 48 intervals between $(u_p)_1=0.58$ and $(u_p)_{48}=0.565$, with $\Delta_E a^4/ 6\tilde V = 0.0006$. The solid blue line shows the final values of $a_n$ after 10 Newton-Raphson iterations followed by 300 Robbins-Monro iterations, with errors calculated from 20 repeats. The black crosses show the measured vacuum expectation value of the average plaquette against $\beta$, calculated from 500,000 configurations generated using conventional importance sampling methods, for $\beta=7.32,7.33,7.34$ and $7.35$. The red dashed line shows the initial $a_n$ values, constructed by fitting a cubic polynomial to the importance sampling results.}
\label{fig:RM}     
\end{figure}

This work focuses on $\SpF$ pure gauge theory on lattice with size $4\times 20 ^3$. 
The energy range is split into 48 intervals, from $(u_p)_1= 0.58$ to $(u_p)_{48}=0.565$, and interval size $\Delta_E a^4 / 6\tilde V = 0.0006$. 
The initial estimates for $a_n$ were based on standard importance sampling results. 
This guess is improved through by 10 Newton-Raphson iterations followed by 300 Robbins-Monro iterations. 
This process has been repeated 20 times to estimate truncation errors. 
The final $a_n$ values and their dependence on the centre of the interval $(u_p)_n$ is shown in blue in Fig.~\ref{fig:RM}. In black is the importance sampling results for $\langle u_p \rangle_\beta$ against $\beta$. The initial guesses for $a_n$ is shown in red. The non-invertible behaviour in the function of $a_n(u_p)$, that is characteristic of a first-order transition, emerges dynamically.

\section{Thermodynamic observables}
\label{sec:results}
The main thermodynamic properties we are interested in studying in this work are the latent heat and the surface tension. 
These observables can be related, through infinite volume and continuum extrapolations, to properties of the plaquette distribution at the critical point, $\beta_c$, at which the two coexisting phases are equally probable --- the plaquette distribution has a double Gaussian structure with two peaks of equal height. 
The difference in the value of plaquette between the two peaks, $\Delta \langle u_p \rangle_{\beta_c}$, can be used to compute the latent heat, Eq.~(33) of Ref.~\cite{lucini2005properties}. 
The logarithm of the ratio of the height of the central (unstable) minima, $P_{min}$, and the degenerate maxima of the plaquette distribution, $P_{max}$,  $-\ln(P_{min}/P_{max})$, can be related to the surface tension, see Eq.~(48) of Ref.~\cite{lucini2005properties}. 
Here we compute these lattice quantities at a single lattice size.

\begin{figure}
     \centering
     \begin{subfigure}[b]{0.49\textwidth}
         \centering
         \includegraphics[width=\textwidth]{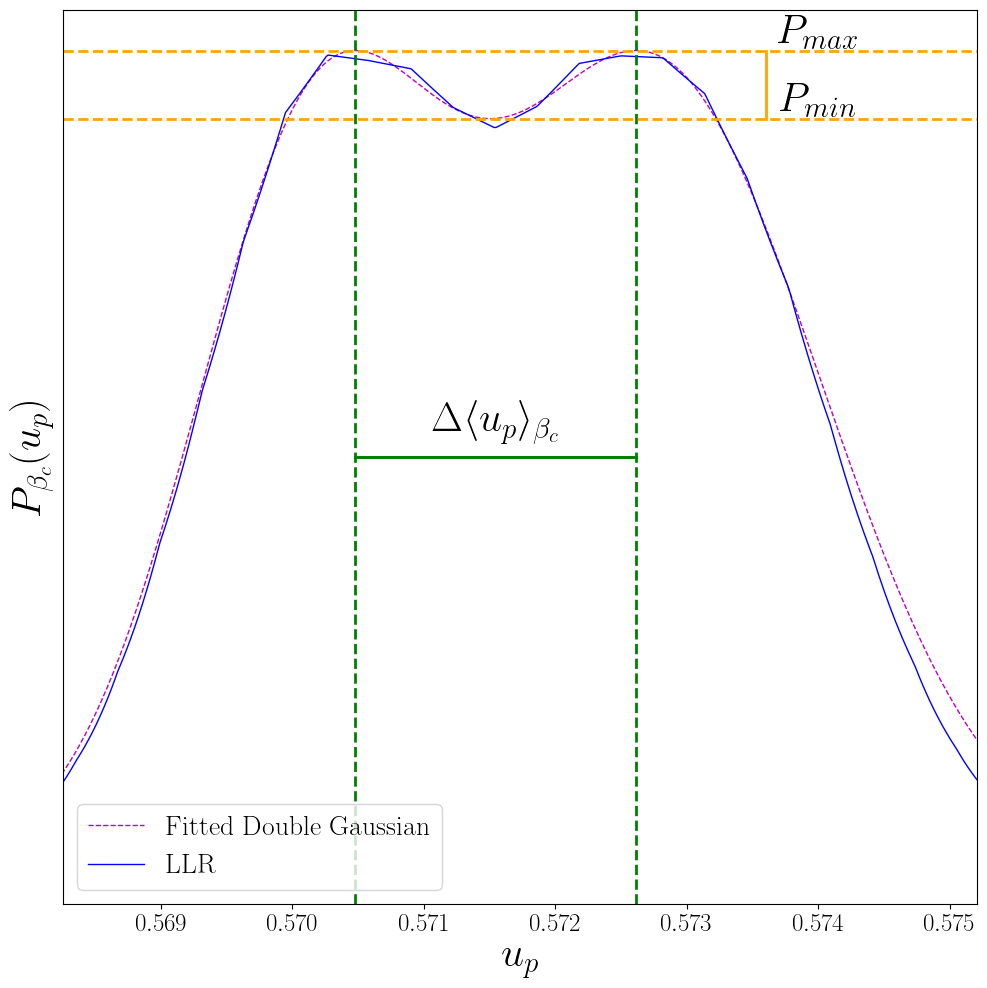}
         \caption{}
         \label{fig:DG}
     \end{subfigure}
     \hfill
     \begin{subfigure}[b]{0.49\textwidth}
         \centering
         \includegraphics[width=\textwidth]{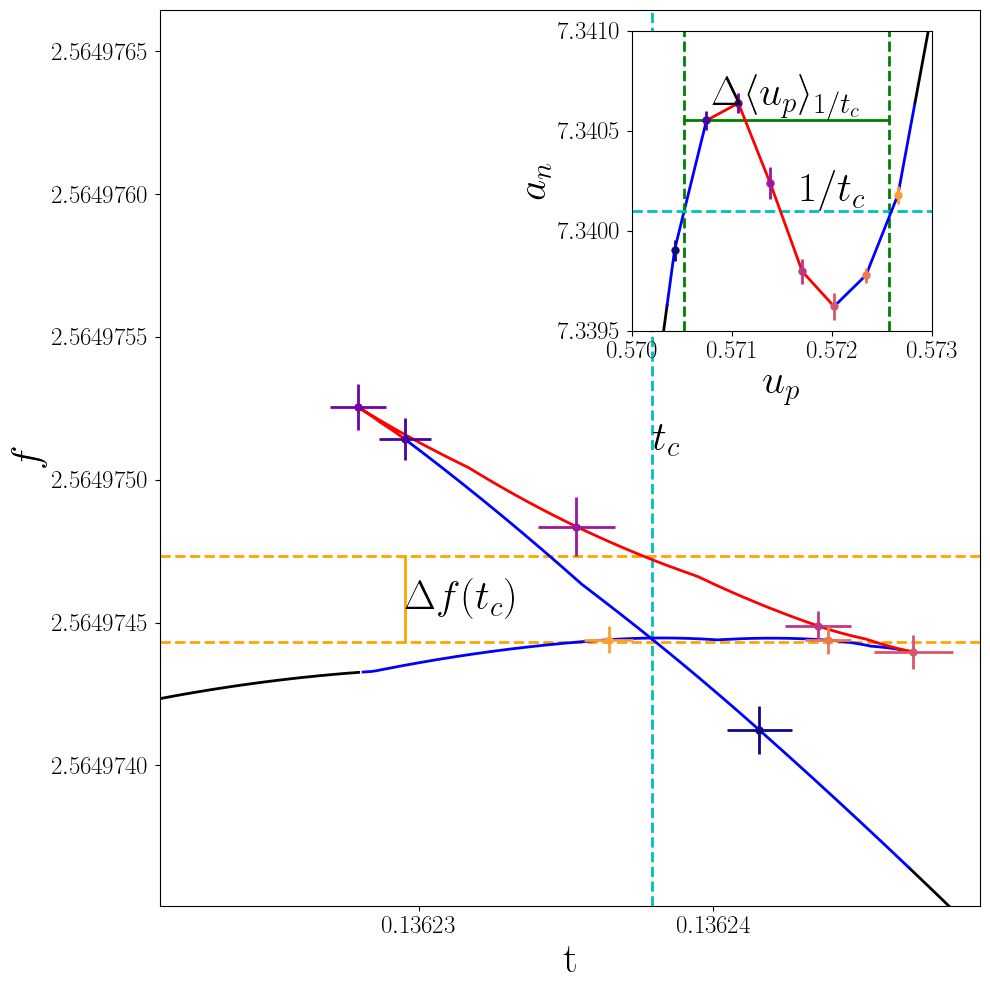}
         \caption{}
         \label{fig:FT}
     \end{subfigure}
        \caption{The reconstructed plaquette distribution at the critical point (left) and the free-energy (right) for $\SpF$ pure gauge theory on a lattice of size $4\times 20^3$, found using the LLR method with 48 intervals between plaquette values of $(u_p)_1=0.58$ and $(u_p)_{48}=0.565$, with $\Delta_E a^4/6 \tilde V = 0.0006$. 
        The critical coupling in the left plot was found by tuning $\beta$ until the two peaks of the distribution have equal height. 
        The plaquette values corresponding to the peaks of the distribution are shown by the green dashed line. 
        The height of the maxima, $P_{max}$, and minima, $P_{min}$, are shown by the orange dashed line. 
        On the right panel the red, blue and black lines show the unstable, metastable and stable regions, respectively. 
        The points in the inset match those of the main plot, showing the corresponding values of $a_n$ and $u_p$. 
        The critical coupling is shown by the dashed cyan line, the orange dashed line shows the free energy values when $t=t_c$ and the green dashed line on the inset show the plaquette values when $a_n=1/t_c$.
        \label{fig:Critical}}
\end{figure}

The critical coupling can be found by fitting a double Gaussian distribution to the plaquette probability distribution, Eq.~(\ref{eqn:Observables}), and tuning the coupling until the difference in height between the two maxima of the distribution is below a certain threshold. 
At this point $\Delta \langle u_p \rangle_{\beta_c}$ and $-\ln(P_{min}/P_{max})$ can be read directly from the distribution, as shown in Fig.~\ref{fig:DG}, with the results presented in Tab.~\ref{tab:lat_surf}. 

The same quantities can also be determined directly from the free energy of the micro-states, $F$, we define as
\beq
\label{eqn:Observables}
F(t)\equiv E-ts, \quad s\equiv\ln \rho \quad t\equiv\frac{\partial E}{\partial s} \equiv \frac{1}{a_n}, \quad f(t) \equiv \frac{a^4}{\tilde V} (F(t) + \Sigma t),
\eeq
where $s$ is the entropy and $t$ is the temperature. $f(t)$ is defined to remove the dependence of the arbitrary $c_1$ term in Eq.~(\ref{eqn:LLRRho}) and make the observed swallow tail structure more clear. The additive constant term $\Sigma$, is equal to the average entropy. 

From the plot of $f$ against $t$, Fig.~\ref{fig:FT}, we can determine the critical point, the plaquette jump and $-ln(P_{min}/P_{max})$. The critical point is the point at which the two metastable branches (solid blue lines) cross (cyan dashed vertical line), with $\beta_c = 1/t_c$. The plaquette jump is determined by the difference between the plaquette values in the metastable regions at which $a_n=\beta_c$ in the plot of $a_n$ against $u_p$, see the inset of Fig.~\ref{fig:FT}. Note we use a linear interpolation between the points.  The free energy is related to the extrema of the energy distribution, through
\beq 
\label{eqn:FreeEnergy}
e^{-\frac{F(t)}{t}} = Z_\beta P_\beta(E)|_{\beta=1/t, E=F(t)+ts}.
\eeq
The change in the free energy between the metastable and unstable branches at the critical point, $\Delta F(t_c)= \tilde V \Delta f(t_c)/a^4$, can be related to the logarithmic term we calculated previously as $-\ln(P_{min}/P_{max})=\tilde V \Delta f(t_c)/a^4t_c$. 

In Tab.~\ref{tab:lat_surf} the results from the probability distribution and from the free energy are presented, for $\SpF$, for a single lattice size. 
They are in good agreement, demonstrating that the methods calculate equivalent values. 
Additionally, included in this table are the results for $\suthree$ for the same lattice size, calculated from the free energy using LLR results presented previously in Ref.~\cite{Lucini:2023irm}. 
Direct comparison of the results, suggest that $\suthree$ has a stronger first-order phase transition as both the plaquette jump is larger and the change in the free energy is larger. Therefore making jump in energy larger and the probability of tunnelling through the barrier lower. 
However, these quantities are not physical. 
Without taking the infinite volume and continuum limits, to gain access to the physical quantities of the continuum theory --- surface tension and latent heat--- we cannot reach any definite conclusions on their relative value.

\begin{table}
\caption{The values of the critical coupling, $\beta_c$, the plaquette jump, $\Delta \langle u_p \rangle_{\beta_c}$, and the change in the free energy divided by the critical temperature, $\Delta F(t_c)/t_c$, are shown for $\suthree$ and $\SpF$  pure gauge theories on $4\times 20^3$ lattices. The $\SpF$ results were found using the LLR method with 48 intervals between plaquette values of $(u_p)_1=0.58$ and $(u_p)_{48}=0.565$, with $\Delta_E a^4/6 \tilde V = 0.0006$. The $\suthree$ values were computed using the results of previous work presented in Ref.~\cite{Lucini:2023irm} with $\Delta_E a^4/6 \tilde V = 0.0007$. The second column notes whether the value was calculated from the free energy,$F(t)$, or from the plaquette distribution,  $P_{\beta_c}(u_p)$.  }
 \label{tab:lat_surf}
\begin{center}
\begin{tabular}{|c|c|c|c|c|}
\hline
  & & $\beta_c = 1/t_c$ & $\Delta \langle u_p \rangle _{\beta_c}$ & $\Delta F(t_c)/t_c = -\ln(P_{min}/P_{max})$\\
\hline
$\suthree$ & $F(t)$  & $5.69189(4)$ & $0.00257(3)$ & $0.0919(83)$ \\
$\SpF$ & $P_{\beta_c}(u_p)$ & $7.34009(3)$ & $0.00203(2)$ & $0.0714(43)$ \\
$\SpF$ & $F(t)$  & $7.34010(3)$ & $0.00205(3)$ & $0.0708(48)$ \\

\hline
\end{tabular}
\end{center}
\end{table}
\section{Conclusion}
We presented initial results for the thermodynamic properties of the deconfinement phase transition in $\SpN$ gauge theories using the LLR method. 
Building on  the methodology presented in Ref.~\cite{Lucini:2023irm}, we calculated the critical coupling, the plaquette jump and the change in the free energy at this coupling for $\SpF$ on a single lattice size $4\times20^3$, presented in Tab.~\ref{tab:lat_surf}. These quantities have been calculated by reconstructing the energy distribution at the critical point, Fig.~\ref{fig:DG}, and separately from the free energy of the micro-states, Fig.~\ref{fig:FT}. 
Both methods are in good agreement. 
These results were compared with results for $\suthree$ at the same lattice size, although without continuum and infinite volume extrapolations these comparisons are not physical. 
In future work, we aim to extend our analysis to additional lattice sizes to accurately compute the physical thermodynamic observables, the latent heat and the surface tension.
\acknowledgments
The work of D.~V. is partly supported by the Simons Foundation under the program ``Targeted Grants to Institutes'' awarded to the Hamilton Mathematics Institute.
The work of D.~M. is supported by a studentship awarded by the Data Intensive Centre for Doctoral Training, which is funded by the STFC grant ST/P006779/1.  
B.~L. and M.~P. received funding from the European Research Council (ERC) under the European Union’s Horizon 2020 research and innovation program under Grant Agreement No.~813942, and by STFC under Consolidated Grants No. ST/P00055X/1, ST/T000813/1, and ST/X000648/1. 
The work of B.~L. is further supported in part by the Royal Society Wolfson Research Merit Award WM170010 and by the Leverhulme Trust Research Fellowship No. RF-2020-4619. 
Numerical simulations have been performed on the Swansea SUNBIRD cluster (part of the Supercomputing Wales project) and AccelerateAI A100 GPU system, on the DiRAC Data Intensive service at Leicester, and on the DiRAC Extreme Scaling service at the University of Edinburgh. The Swansea SUNBIRD system and AccelerateAI are part funded by the European Regional Development Fund (ERDF) via Welsh Government. 
The DiRAC Data Intensive service at Leicester is operated by the University of Leicester IT Services, and the DiRAC Extreme Scaling service is operated by the Edinburgh Parallel Computing Centre, they form part of the STFC DiRAC HPC Facility (www.dirac.ac.uk). 
The DiRAC Data Intensive service equipment at Leicester was funded by BEIS capital funding via STFC capital grants ST/K000373/1 and ST/R002363/1 and STFC DiRAC Operations grant ST/R001014/1. 
The DiRAC Extreme Scaling service was funded by BEIS capital funding via STFC capital grant ST/R00238X/1 and STFC DiRAC Operations grant ST/R001006/1.
DiRAC is part of the National e-Infrastructure.\\

{\bf Open Access Statement - } For the purpose of open access, the authors have applied a Creative Commons 
Attribution (CC BY) licence  to any Author Accepted Manuscript version arising.

{\bf Research Data Access Statement}---The results for $\SpF$ are based on preliminary analysis. Further analysis and the data generated for this manuscript will be released together with an upcoming publication. The simulation code and data generated for the \suthree results can be found in data releases Ref.~\cite{LMPRV2} and Ref.~\cite{LMPRV} respectively.  
\bibliographystyle{jhep}
\bibliography{references}
\end{document}